\documentclass[journal]{IEEEtran}
\usepackage{amsfonts,amssymb}
\usepackage{amsmath}
\usepackage{amsthm}

\usepackage{graphicx}
\usepackage{enumerate}
\usepackage{multirow}
\usepackage{longtable}
\usepackage{rotating}
\usepackage{calc}
\usepackage{subfig}
\usepackage{graphicx}
\usepackage{epstopdf}
\usepackage{graphics}
\usepackage{epsfig}
\usepackage{psfrag}
\usepackage{cases}

\usepackage{array,color}
\usepackage{tabularx}
\usepackage{multirow}
\usepackage{booktabs}
\usepackage{makecell}  

\usepackage{algorithm}

\usepackage{algorithmic}

\usepackage{multirow}

\usepackage[numbers,sort&compress]{natbib}
\usepackage{url}
\usepackage{bm}
\usepackage{amsfonts}
\usepackage{amsmath}
\usepackage{amssymb}
\usepackage{amsthm}
\usepackage{color}

\usepackage{array,color}
\usepackage{tabularx}
\usepackage{multirow}
\usepackage{booktabs}
\usepackage{makecell}  

\usepackage{graphicx}  
\usepackage{subfig}
\usepackage{epstopdf}
\usepackage{graphics}
\usepackage{epsfig}
\usepackage{psfrag}
\usepackage{overpic}

 \usepackage{stfloats}
 \usepackage{float}   

 \usepackage{indentfirst} 

\usepackage{gensymb}

\begin{document}

\title{A Novel Approach for Big Data Analytics in Future Grids Based on Free Probability}

\author{Zenan Ling,   Robert~C. Qiu,~\IEEEmembership{Fellow,~IEEE}, Xing He,    and Chu Lei
\thanks{This work was partly supported by National Natural Science Foundation of China (No. 51577115 and No. 61571296).}
}

\maketitle

\IEEEpeerreviewmaketitle
\begin{abstract}
Based on the random matrix model, we can build statistical models using massive datasets across the power grid, and employ hypothesis testing for anomaly detection. First, the aim of this paper is to make the first attempt to apply the recent free probability result in extracting big data analytics, in particular data fusion. The nature of this work is basic in that new algorithms and analytics tools are proposed to pave the way for the future's research. Second,  using the new analytic tool, we are able to make some discovery related to anomaly detection that is very difficult for other approaches. To our best knowledge, there is no similar report in the literature. Third, both linear and nonlinear polynomials of large random matrices can be handled in this new framework. Simulations demonstrate the  following: Compared with the linearity, nonlinearity is more flexible in problem modeling and closer to the nature of the reality. In some sense, some other nonlinear matrix polynomials may be more effective for the power grid.

\end{abstract}


\section{Introduction}

\IEEEPARstart{D}{ata}-driven approach and data utilization for smart grids are  current stressing topics, as evidenced in the special issue of  ``Big Data Analytics for Grid Modernization''~\cite{bda2016tsg}. Among some challenges, data fusion is of great significance in power system operation in future grids~\cite{bda2016tsg}, which are always huge in size and  complex in topology.

Randomness or uncertainty is at the heart of this data modeling and analysis. Data fusion brings together the massive datasets across the power grid. Due to the data size, randomness plays in a basic role in extracting big data analytics.  Our approach exploits the massive datasets across the grid that are distributed in both spatially and temporally. Random matrix theory (RMT) appears very natural for the problem at hands; in a random matrix of  ${\mathbb{C}^{N \times T}},$ we use $N$ nodes to represent the spatial nodes and $T$ data samples to represent the temporal samples.

When the number of nodes $N$ is large, very unique mathematical phenomenon occurs, such as free probability~\cite{qiu2015smart}. Phase transition as a function of data size $N$ is a result of this deep mathematical phenomenon. This is the very reason why the proposed algorithms are so powerful in practice. Asymptotic limits of $N\to \infty$ and $T\to \infty$ can be obtained through algorithms, although closed form expressions exist only for some simple matrix polynomials.

This paper is built upon our previous work in the last several years. See Section~\ref{sect:RelationshipPrevious} for details. Motivated for data mining, our line of research is based on the high-dimensional statistics. By high-dimensionality, we mean that the datasets are represented in terms of large random matrices. These data matrices can be viewed as data points in high-dimensional vector space of mathematics---each vector is very long.

Randomness is critical to a complex, large power grid in the future since rapid fluctuations in voltages and currents are ubiquitous. Often, these fluctuations exhibit Gaussian statistical properties~\cite {lim2016svd}. Our central interest in this paper is to model these rapid fluctuations, using the framework of random matrix theory.  Our new algorithms are made possible due to the \textit{latest breakthroughs} in free probability.

\subsection{Relationship to the Prior Art}
\label{sect:RelationshipPrevious}

Unified by large random matrices through RMT, our sequence of three monographs explores big data analytics in wireless network~\cite{qiu2012bookcogpp}, sensing~\cite{qiu2013bookcogsen} and smart grid~\cite{qiu2015smart}. In~\cite{zhang2015MassiveMIMO,Zhang2014Data,he2016big}, large random matrices are used with RMT to model experimental data in a large wireless communication network. Following this success, our work~\cite{he2015arch}  is the first attempt to introduce the mathematical tool of RMT into power systems. Later, numerous papers \cite{he2015corr} demonstrate the power of this tool. Ring Law and Marchenko-Pastur (M-P) Law are regarded as the statistical foundation, and Mean Spectral Radius (MSR) is proposed as the high-dimensional indicator.

Then we move forward to the second stage---paper \cite{he2015corr} studies the correlation analysis under the above framework. The concatenated matrix $\mathbf{A}_i$ is the object of interest. It consists of the basic matrix $\mathbf{B}$ and a factor matrix $\mathbf{C}_i$, i.e., $\mathbf{A}_i\!=\![\mathbf{B}; \mathbf{C}_i]$. In order to seek the sensitive factors, we compute the  advanced indicators that are based on  the linear eigenvalue statistics~\cite{Qiu2016BigDataLRM,he2015les} of these concatenated matrices $\mathbf{A}_i$. This study contributes to fault detection and location, line-loss reduction, and power-stealing prevention \cite{xu2016Corr}.
We also conduct analysis for power transmission equipment based on the same theoretical foundation~\cite{yan2016transmission}.

The success of using the concatenated matrix $\![\mathbf{B}; \mathbf{C}_i]$ to ``combine'' the data matrices $\bf B$ and ${\bf C}_i$ has inspired us to explore alternative techniques for such a purpose, called data fusion.

\subsection{Contributions of Our Paper}

Our prior work is based on the sample covariance matrix $S,$  which a Wishart matrix when all the entries of $X$ of $N\times T$ are Gaussian random variables. As pointed above, this model appears suitable.

Let us consider two sample covariance matrices $S_0$ and $S_1,$ that are collected in different spatial (and/or temporal) parts of the power grid. We ask this basic question: What are the natural manners for combining the matrices $S_0$ and $S_1$? Answer: (1)  linear matrix polynomial ${P_1}({S_0},{S_1}) = {S_0} + {S_1};$ (2) self-adjoint matrix nonlinear polynomial:  ${P_2}({S_0},{S_1}) = {S_0}{S_1} + {S_1}{S_0}.$

It is very difficult to obtain solutions for the above two cases. Fortunately, the recent breakthrough in free probability in random matrix theory has made this possible. The advanced tool~\cite{Speicher2015Polynomials} is highly inaccessible to the power field.

\begin{enumerate}
\item  The aim of this paper is to make the first attempt to apply the recent free probability result in extracting big data analytics, in particular data fusion. The nature of this work is basic in that new algorithms and analytics tools are proposed to pave the way for the future's research.
	\item Using the new analytic tool, we are able to make some discovery related to anomaly detection that is very difficult for other approaches. To our best knowledge, there is no similar report in the literature.
	\item  Both linear and nonlinear polynomials of large random matrices can be handled in this new framework. Simulations demonstrate the  following: Compared with the linearity, nonlinearity is more flexible in problem modeling and closer to the nature of the reality. In some sense, some other nonlinear matrix polynomials may be more effective for the power grid.
	\end{enumerate}

\section{Data-driven modeling for power grid}
\subsection{Random Matrix Model For Power Grid}
Following \cite{He2016Designing}, we build the statistic model for power grid.
Considering   $T$ random vectors observed at time instants $i=1,...,T,$ we form a random matrix as follows
 \begin{equation}
\label{eq:RMTform}
	\left[ {\Delta {{{\mathbf{V}}_1}} , \cdots ,\Delta  {{{\mathbf{V}}_T}}} \right] = \left[ { {{\mathbf{\Xi}}_1\Delta{\mathbf{P}}_1}, \cdots , {{{\mathbf{\Xi}}_T\Delta \mathbf{P}}_T}} \right].
\end{equation}

In an equilibrium operating system, the voltage magnitude vector injections ${\bf V}$ with entries $V_{i},i=1,\cdots,N$ and the phase angle vector injections $\boldsymbol{\theta}$ with entries $\theta_{i},i=1,\cdots,N$ remain relatively constant. Without dramatic topology changes, rich statistical empirical evidence indicates that the Jacobian matrix $\mathbf{J}$  keeps nearly constant, so does $\mathbf{\Xi}$. Also, we can estimate the changes of ${\bf V},$ $\boldsymbol{\theta},$ and $\mathbf{\Xi}$ only with the \emph{empirical} approach.
Thus  we rewrite \eqref{eq:RMTform} as:

\begin{equation}
\label{Eq:RMMVTP}
\mathbb{V} = {\bm{\Xi}_N}{\mathbb{P}}_{N \times T}
\end{equation}
where $\mathbb{V} = \left[ {\Delta {{{\mathbf{V}}_1}} , \cdots ,\Delta  {{{\mathbf{V}}_T}}} \right]$, $ {\bm {\Xi}}={\mathbf{\Xi}}_1=\cdots={\mathbf{\Xi}}_T,$ and $\mathbb{P} = \left[ {\Delta {{{\mathbf{P}}_1}} , \cdots ,\Delta  {{{\mathbf{P}}_T}}} \right].$
Here $\mathbb{V}$ and $\mathbb{P}$ are random matrices. In particular, $\mathbb{P}$ is a random matrix with Gaussian random variables as its entries.

\newtheorem{lemma}{\textbf{Lemma}}[section]

 \begin{lemma}[M-P Law~\cite{qiu2012bookcogpp,qiu2013bookcogsen,qiu2015smart}]

 Let $X = \{ {x_{i,j}}\} $ be a $N \times T$ random matrix whose entries with the mean $\mu= 0$ and the variance ${\sigma ^2}<\infty $, are independent identically distributed (i.i.d). As $N, T \longrightarrow \infty$ with the ratio $ c=N/T \in (0,1] $.
 \begin{equation}
\label{defeq:SampleCov}
S = \frac{1}{N}X{X^H} \in {{\mathbb{C}}^{N \times N}}	
 \end{equation}
	is the  corresponding sample covariance matrix. Then,  the asymptotic spectral distribution of $S$ is given by:
 \begin{equation}
  {\mu ^{' }}(x)=
  \begin{cases}
  \frac{1}{{2\pi x{\sigma ^2}}}\sqrt {(b - x)(x - b)} &\mbox{if $a\leq x\leq b$}\\
  0 &\mbox{otherwise}
  \end{cases}
 \end{equation}
 where $a = {\sigma ^2}{(1 - \sqrt c )^2}$, $b = {\sigma ^2}{(1 + \sqrt c )^2}$. Here, $S$ is called Wishart matrix.
 \label{lem1}
\end{lemma}

\section{Data Fusion Method}
 Based on  the  random matrix model proposed in \cite{He2016Designing,He2015A}, we can build statistical models only using the sampling data and employ hypothesis testing
 for anomaly detection. In the situation of one random matrix, the spectral distribution of the sample covariance matrix obeys M-P Law if it is Wishart matrix according to Lemma \ref{lem1}.
 Therefore, hypothesis testing for anomaly detection is conducted by comparing the  spectral distribution of sample covariance matrices with M-P Law. This anomaly detection method performs well
 in detecting step signals. However, it is not effective when signals are changing  continuously and slowly (see  Section~\ref{case1}). In most cases, anomaly detection methods with data fusion performs
 better than that without data fusion \cite{Siaterlis2011Theory}. Therefore, we are now interested in data fusion methods based on random  matrix models for the power grid.

\subsection{Data Fusion Models }
\subsubsection{Model designs}
Multivariate  linear or nonlinear polynomials perform a significant role in problem modeling, so we build our data fusion models on the basis of random matrix polynomials.
  In this paper, we study two typical random matrix polynomial models.

 The first one is the multivariate linear polynomial: \[{P_1}({S_0},{S_1}) = {S_0} + {S_1}.\]

The second one is the self-adjoint multivariate nonlinear polynomial:  \[{P_2}({S_0},{S_1}) = {S_0}{S_1} + {S_1}{S_0}.\]

Here, both ${S_0}$ and ${S_1}$ are the sample covariance matrices.

\subsection{Theoretical Bound}
 In this subsection, we study  the asymptotic spectral distribution (ASD) of ${P_i}$, $i=1,2$, on the premise that both ${S_0}$ and ${S_1}$ are Wishart matrices.

\subsubsection{The ASD of ${P_1}$}

We obtain the ASD of ${P_1}$ based on the operator-valued setting in ~\cite{Speicher2015Polynomials}. The brief introduction of this setting is in the appendix.
Stieltjes-Cauchy transform is an effective tool in free probability theory.
\newtheorem{mydef}{\textbf{Definition}}[section]
\begin{mydef}[Stieltjes-Cauchy transform]
Consider a non-negative and finite Borel measure $\mu$ on $\mathbb{R}$. For all $z \in \{z:z \in \mathbb{C}, \Im (z)>0\}$, the Stieltjes transform of $\mu$ is defined as

\[{G_\mu }(z) = \int\limits_\mathbb{R} {\frac{1}{{z - x}}} d\mu (x).\]

\end{mydef}

The definition of the operator-valued Cauchy transform is necessary for Theorem~\ref{th1}

 \newtheorem{def2}{\textbf{Definition}}[section]
 \begin{mydef}[the operator-valued Cauchy transform]

Let $\mathcal{A}$ be a unital algebra and $\mathcal{B}\subset \mathcal{A}$ be a subalgebra containing the unit. A linear map $E:\mathcal{A}\to \mathcal{B}$ is a conditional expectation.
For a random variable $x\in \mathcal{A}$ ,  the operator-valued Cauchy transform is defined as $G(b):=E[{{(b-x)}^{-1}}]   (b\in \mathcal{B})$ for which $(b-x)$ is invertible in $\mathcal{B}$ .

 \end{mydef}

\newtheorem{theorem}{\textbf{Theorem}}[section]
 \begin{theorem}[\cite{Belinschi2013Analytic}]
 Let $x$  and $y$ be selfadjoint operator-valued random variables free over $\mathcal{B}$. Then there exists a Frechet analytic map
$\omega :{{\mathbb{H}}^{+}}(\mathcal{B} )\to {{\mathbb{H}}^{+}}(\mathcal{B})$ so that

$\bullet\Im {{\omega }_{j}}(b)\ge \Im b$ for all  $b\in {{\mathbb{H} }^{+}}(\mathcal{B} )$, $j\in \left\{ 1,2 \right\}$

$\bullet{{G}_{x}}({{\omega }_{1}}(b))={{G}_{y}}({{\omega }_{2}}(b))={{G}_{x+y}}(b)$

Moreover, if $b\in {{\mathbb{H}}^{+}}(\mathcal{B})$ , then ${{\omega }_{1}}(b)$  is the unique fixed point of the map.
${{f}_{b}}:{{\mathbb{H}}^{+}}(\mathcal{B} )\to {{\mathbb{H}}^{+}}(\mathcal{B} ),  {{f}_{b}}(\omega )={{h}_{y}}({{h}_{x}}(\omega )+b)+b, $
 and ${{\omega }_{1}}(b)\text{=}\underset{n\to \infty }{\mathop{\lim }}\,{{f}_{b}}^{on}(\omega )$ for any $\omega \in {{\mathbb{H}}^{+}}(\mathcal{B} )$, where $f_{b}^{on}$ means the n-fold composition of ${{f}_{b}}$ with itself. Same statements hold for ${{\omega }_{\text{2}}}(b)$, with replaced by $\omega \to {{h}_{x}}({{h}_{y}}(\omega )+b)+b. $
  \label{th1}
  \end{theorem}

\newtheorem{xxxx}{\textbf{Theorem}}[section]
 \begin{theorem}[Stieltjes inversion formula \cite{Pielaszkiewicz2015Closed}]
 For any open interval $I=(a,b)$ , such that neither a nor b are atoms for the probability measure $\mu $, the inversion formula
	\[\mu (I)=-\frac{1}{\pi }\int\limits_{I}{\Im ({{G}_{u}}(x+iy))dx}\]
holds.

 \end{theorem}

By Theorem \ref{th1}, we  achieve the operator-valued Cauchy transform of ${P_1}$. Then, the ASD  of ${P_1}$ is easily obtained through the Stieltjes inversion formula.

\subsubsection{The ASD Of ${P_2}$}
  The ASD of ${P_2}$ is obtained  by linearizing the nonlinear polynomial and   Theorem \ref{th1}.  Through Anderson's  linearation trick \cite{Anderson2011Convergence}, we have a
 procedure that leads finally to an operator:
  \[{{L}_{p}}={{b}_{0}}\otimes 1+{{b}_{1}}\otimes {{S}_{1}}+\cdots {{b}_{n}}\otimes {{S}_{n}}.\]
 In the case of ${P_2}$,  \[{L_{{P_2}}} = \left( {\begin{array}{*{20}{c}}
0&{{S_1}}&{{S_2}}\\
{{S_1}}&0&{ - 1}\\
{{S_2}}&{ - 1}&0
\end{array}} \right)\]

Therefore, ${L_{{P_2}}}$ can be easily written in the form of ${L_{{P_2}}}={{b}_{0}}\otimes 1+{{b}_{1}}\otimes {{S}_{1}}+ {{b}_{2}}\otimes {{S}_{2}}$, where ${b_0} = \left( {\begin{array}{*{20}{c}}
0&0&0\\
0&0&{ - 1}\\
0&{ - 1}&0
\end{array}} \right),$
${b_1} = \left( {\begin{array}{*{20}{c}}
0&1&0\\
1&0&0\\
0&0&0
\end{array}} \right),$
${b_2} = \left( {\begin{array}{*{20}{c}}
0&0&1\\
0&0&0\\
1&0&0
\end{array}} \right).$

Then, we obtain the algorithm to effectively compute the ASD of ${P_2}$ by applying Theorem \ref{th1} and Stieltjes inversion formula. See the detail derivation and specific procedures of the algorithm in Section~\ref{sect:AlgorithmFreeAdjointPolynomial}.

\subsection{The Processing of the Grid Data}
\label{C}
The sampling data of power grid is always non-Gaussian, so  a  normalization procedure in \cite{He2015A} is adopted to conduct data preprocessing. Meanwhile,  a Monte Carlo method is employed to compute the spectral distribution of raw data polynomial according to the asymptotic property theory. The details  are in the following algorithm.

\begin{algorithm}[h]

    \caption{}
    \label{alg1}
    \begin{algorithmic}[1]
\REQUIRE ~~\\
The sample data matrices: ${V_0}$ and ${V_1}$ ;\\
The number of repetition times : $M$;\\
The size of ${V_0}$ and ${V_1}$: $N,T$;\\
SNR: $\eta$
        \FOR{$i \leq M$}

            \STATE add  small white noises to sample data matrices；\\
            $ \widetilde{V_0}= V_0 + \eta$ randn($N,N$);\\
            $ \widetilde{V_1}= V_0 + \eta$ randn($N,N$);
            \STATE standardize $\widetilde{V_0}$ and $\widetilde{V_1}$, i.e. mean=0, variance=1;
            \STATE generate the covariance matrices: ${S_0}=\widetilde{V_0}\widetilde{V_0}^{'}/N$,\\
             ${S_0}=\widetilde{V_1}\widetilde{V_0}^{'}/N$;
            \STATE compute the eigenvalues of $P({S_0},{S_1})$;
        \ENDFOR
\STATE Computer the frequency of different eigenvalues and draw the spectral distribution histogram;
\ENSURE ~~\\
The spectral distribution  histogram.

    \end{algorithmic}

\end{algorithm}

\subsection{Hypothesis Testing and Anomaly Detection With Data fusion}

We formulate our problem of anomaly detection in terms of the same hypothesis testing as \cite{He2016Designing}: no outlier exists ${\cal H}_0$, and outlier exists ${\cal H}_1$.

\begin{equation}
\left|
\begin{array}{*{20}{c}}
{{{\cal H}_0}:\widetilde{\mathbb{V}}  = \widetilde{\Xi}}{R}_{N\times T}\\
{{{\cal H}_1}:\widetilde{\mathbb{V}}  \neq \widetilde{\Xi}}{R}_{N\times T}
\end{array}
\right.
\end{equation}
where ${R}$ is the standard Gaussian random matrix.

Our detection method is summarized as follows. Firstly, generate ${S_0}$, ${S_1}$ from the sample data through the preprocess in \ref{C}. Then, compare the theoretical bound with the spectral distribution of raw data polynomials.
If outlier exists, ${\cal H}_0$ will be rejected, i.e. signals exist in the system.

\section{ The Algorithm for Free Adjoint Polynomial }
\label{sect:AlgorithmFreeAdjointPolynomial}
       \subsection{From the Linearization $P_2$ of to the Cauchy Transform of $P_2$}

Through Anderson's linearization trick, we linearize $P_2$:  \[{{L}_{P_2}}={{b}_{0}}\otimes 1+{{b}_{1}}\otimes {{S}_{1}}+ {{b}_{1}}\otimes {{S}_{2}}.\]
To make sure that  $z-{P_2}$  is invertible in $\mathcal{A}$ if and only if $\Lambda (z)-{{L}_{P_2}}$ is invertible in ${{M}_{N}}(\mathbb{C})\otimes \mathcal{A}$,  we let \[\Lambda (z):=\left[ \begin{matrix}
   z & 0 & \cdots  & 0  \\
   0 & 0 & \cdots  & 0  \\
   \vdots  & \vdots  & \ddots  & \vdots   \\
   0 & 0 & \cdots  & 0  \\
\end{matrix} \right]     \quad    for \;all\; {z}\in {\mathbb{C}}.\]
The the operator-valued Cauchy transform of $P_2$ is easily obtained. However, the Cauchy transform of $P_2$ is our desired result. The following corollary helps us to recover the Cauchy transform of $P_2$ from the the operator-valued Cauchy transform of $P_2$.

\newtheorem{corollary}{\textbf{Corollary}}[section]
\begin{corollary}[\cite{Belinschi2013Analytic}]
 Consider that $p\in \mathbb{C}<{{X}_{1}},\ldots ,{{X}_{n}}>$ that has a selfadjoint linearization\[{{L}_{p}}={{b}_{0}}\otimes 1+{{b}_{1}}\otimes {{X}_{1}}+\cdots {{b}_{n}}\otimes {{X}_{n}}\]
Then, for each $z\in {{\mathbb{C}}^{+}}$ and all  small enough $\varepsilon >0$ , the operators $z-P\in \mathcal{A}$ and ${{\Lambda }_{\varepsilon }}(z)-{{L}_{P}}\in {{M}_{N}}(\mathbb{C})\otimes \mathcal{A}$ are both invertible and  \[{{G}_{P}}(z)\text{=}\underset{\varepsilon \to \text{0}}{\mathop{\lim }}\,{{\left[ G{}_{{{L}_{P}}}({{\Lambda }_{\varepsilon }}(z)) \right]}_{1,1}}\quad for\; all \;z\in {{\mathbb{C}}^{+}}\] holds.

\label{co1}
\end{corollary}

Then,the ASD of $P_2$ is finally obtained through the Stieltjes inversion formula.

\subsection{Algorithm}
We conclude the algorithm as following steps.

\begin{enumerate}[step 1]

\item Compute the linearization of $P_2$
\[{{L}_{P_2}}={{b}_{0}}\otimes 1+{{b}_{1}}\otimes {{S}_{1}}+ {{b}_{2}}\otimes {{S}_{2}}\] through Anderson's linearization trick.

\item Compute the  Cauchy transform ${{G}_{{{b}_{j}}\otimes {{S}_{j}}}}(b)$ through the scalar-valued Cauchy transforms :
\[{{G}_{{{b}_{j}}\otimes {{S}_{j}}}}(b)=\underset{\varepsilon \to 0}{\mathop{\lim }}\,-\frac{1}{\pi }\int_{\mathbb{R}}{(b-t{{b}_{j}}}{{)}^{-1}}\Im ({{G}_{{{S}_{j}}}}(t+i\varepsilon ))dt.\]
for $j=1,2$.
\item Calculate the Cauchy transform of
\[{{L}_{P_2}}-{{b}_{0}}\otimes 1={{b}_{1}}\otimes {{S}_{1}}+ {{b}_{2}}\otimes {{S}_{n}}\]
by applying Theorem \ref{th1}.
The Cauchy transform of ${{L}_{P_2}}$ is then given by \[{{G}_{{{L}_{P_2}}}}(b)={{G}_{{{L}_{P_2}}-{{b}_{0}}\otimes 1}}(b-{{b}_{0}}).\]

\item According to  Corollary  \ref{co1}, the scalar-valued Cauchy transform ${{G}_{P_2}}(z)$ of $P_2$ 	is obtained by
\[{{G}_{P_2}}(z)\text{=}\underset{\varepsilon \to \text{0}}{\mathop{\lim }}\,{{\left[ G{}_{{{L}_{P_2}}}({{\Lambda }_{\varepsilon }}(z)) \right]}_{1,1}}\quad for\; all \;z\in {{\mathbb{C}}^{+}}.\]

\item Compute the distribution of $P_2$ via the Stieltjes inversion formula.

\end{enumerate}

\section{CASE STUDIES}

\begin{figure}[htpb]
\centering
\begin{overpic}[scale=0.58
]{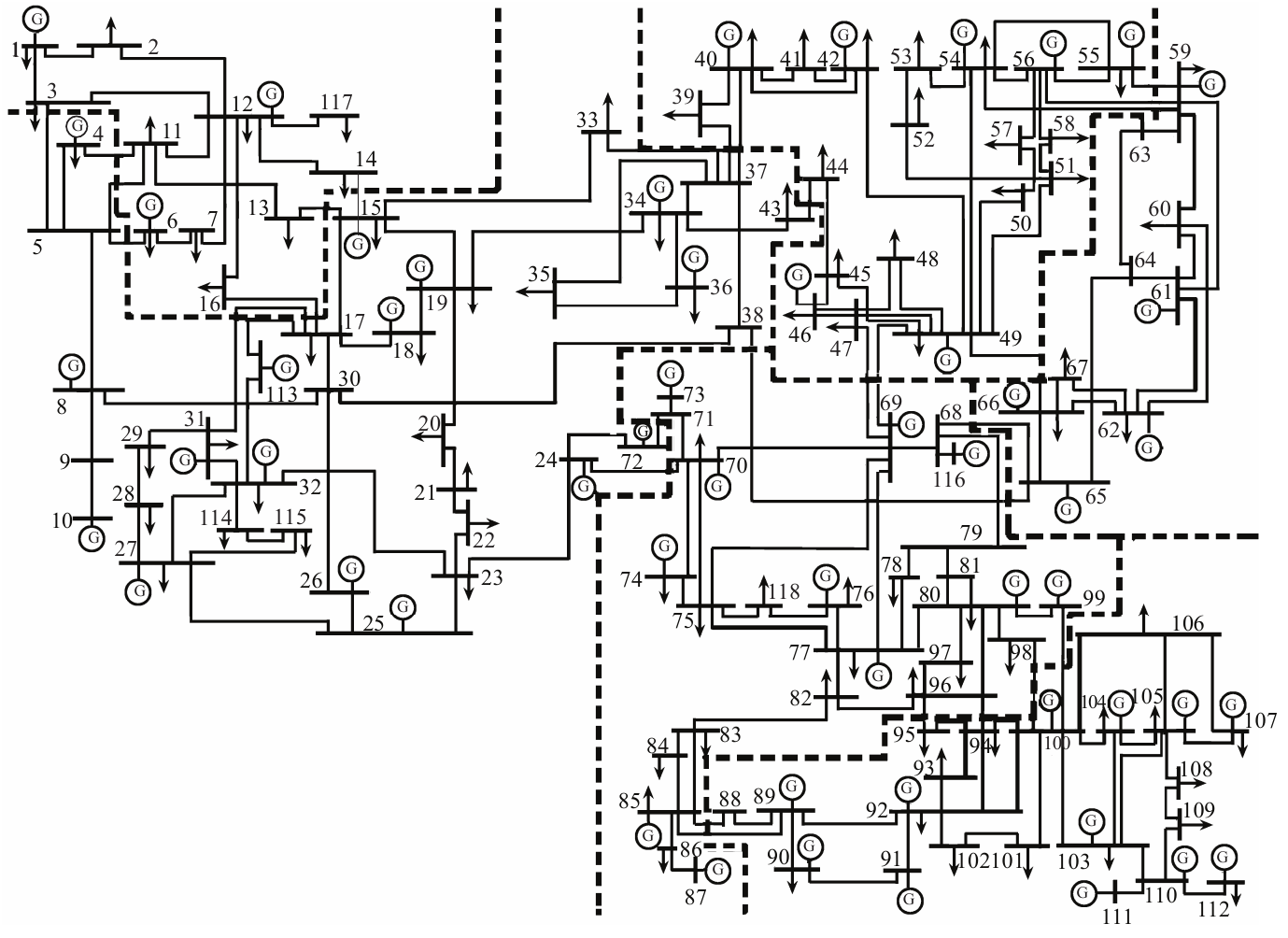}

\end{overpic}
\caption{Partitioning network for the IEEE 118-bus system. There are six partitions, i.e. A1, A2, A3, A4, A5, and A6.}
\label{fig:IEEE118network}
\end{figure}

Our data fusion method is tested with simulated data in the standard IEEE 118-bus system, as shown in Fig. \ref{fig:IEEE118network}. Detailed information of the system is referred to the case118.m in Matpower package and Matpower 4.1 User's Manual \cite{Zimmerman2011MATPOWER}. For all cases, let the sample dimension $N=118$.

\begin{figure*}[htpb]
\centering
\begin{overpic}[scale=0.85
]{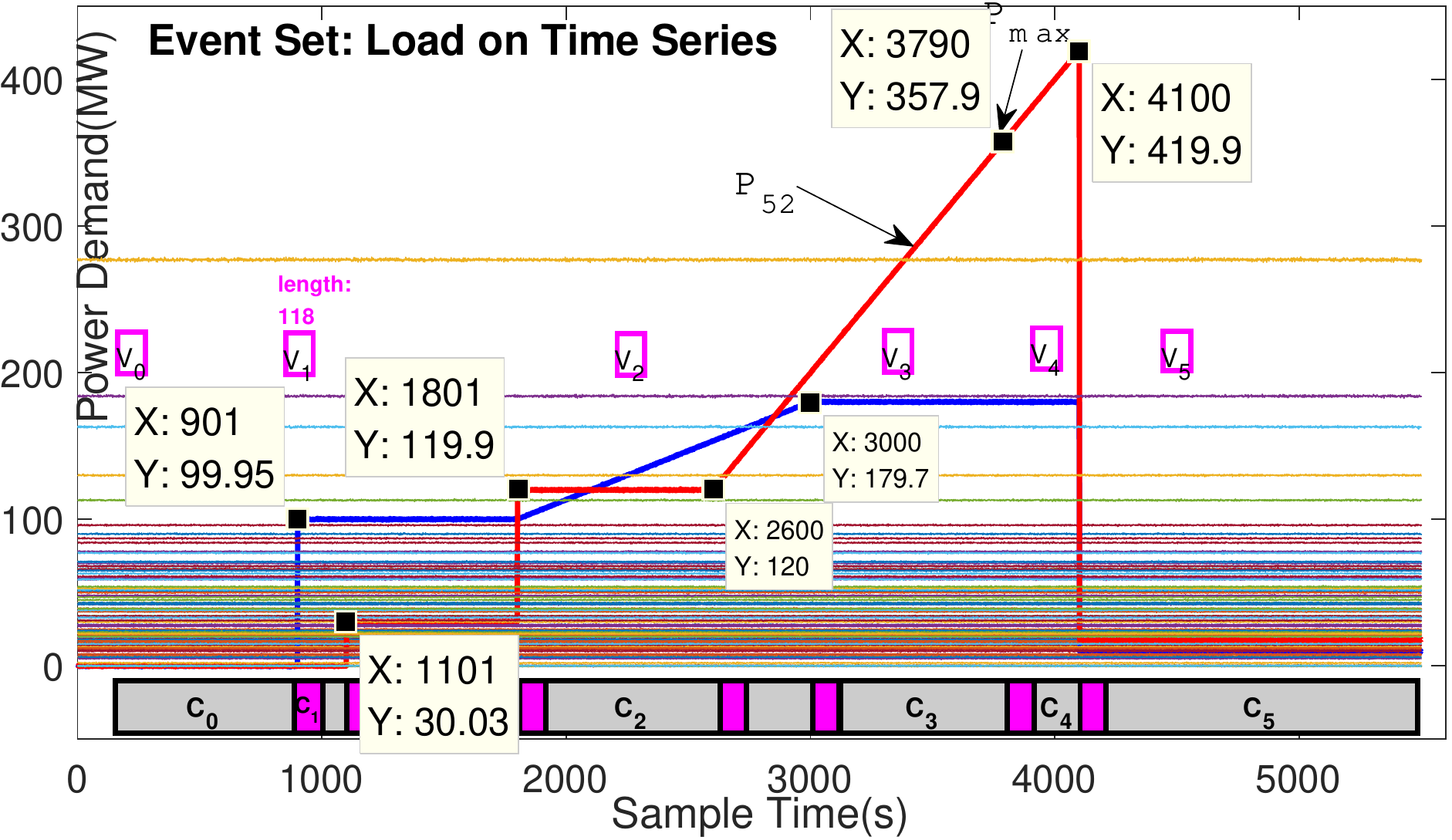}

\end{overpic}
\caption{The event assumptions on time series.}
\label{fig:loadevent}
\end{figure*}

The spectral density distribution of free adjoint polynomial can be obtained through the algorithm in  Section~\ref{sect:AlgorithmFreeAdjointPolynomial} as long as $T \le N$. Meanwhile, the sample dimension $N$ is large enough to guarantee the accuracy of results due to the asymptotic property theory. Therefore, in our simulations, we set the sample length equal to $N$, i.e. $T=118$, $c=T/N=1$ and select six sample voltage matrices presented in Tab. \ref{tab1}, as shown in Fig. \ref{fig:loadevent}.

\begin{table}[h]

\centering

\caption{System status and sampling data}

\begin{tabular}{p{2.05cm}|p{2.2cm}|p{3.58cm}}
\hline
  Cross Section (s)& Sampling (s) &Descripiton\\
\hline
  $\textbf{C}_0:118-900$ & $V_0:100\sim217$&Reference, no signal\\
  $\textbf{C}_1:901-1017$ & $V_1:850 \sim 967$&Existence of a step signal \\
  $\textbf{C}_2:1918-2600$ & $V_2:2200 \sim 2317$&Steady load growth for Bus 22\\
  $\textbf{C}_3:3118-3790$ & $V_3:3300 \sim 3417$&Steady load growth for Bus 52\\
  $\textbf{C}_4:3908-4100$ & $V_4:3900 \sim 4017$&Chaos due to voltage collapse\\
  $\textbf{C}_5:4118-5500$ & $V_5:4400 \sim 4517$&No signal\\
\hline
\end{tabular}
\label{tab1}
\raggedright
 {*We choose the temporal end edge of the sampling matrix as the marked time for the cross section. E.g., for $V_0:100\sim217$, the temporal label is 217 which belong to $\textbf{C}_0:118-800$. Thus, this method is able to be applied to conduct real-time analysis.}
 \end{table}

 Power grid operates with only white noises during 0 s to 900 s; we choose sampling matrix {$V_0$}  as  the reference. Similarly, we mark other kinds of system operation status as  $\textbf{C}_1$--$\textbf{C}_5$, and choose their relevant sampling matrix {$V_1$}--{$V_5$} for the test.

\subsection{Case 1: Detection without data fusion}
\label{case1}
We would like to refer to  the simulation in \cite{He2015A}  as a clue by comparing  the spectral distribution histogram of the matrix polynomial ${P_0}$ with M-P Law. Here, \[{P_0}({S_i}) = {S_i},i=0,1,2,3,4,5.\]

Simulation results are shown in Fig.3.

\begin{figure*}[htpb]
 \centering
 \subfloat[Reference $V_0$]{\label{fig1a}
 \includegraphics[width=0.3\textwidth]{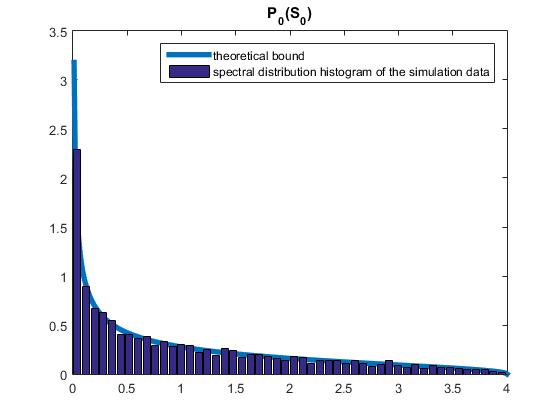}
 }

 \subfloat[Step signal $V_1$]{\label{fig1b}
 \includegraphics[width=0.3\textwidth]{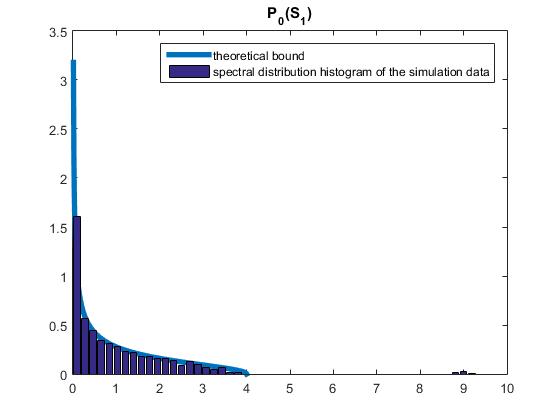}
 }
 \subfloat[Stable growth A $V_2$]{\label{fig1c}
 \includegraphics[width=0.3\textwidth]{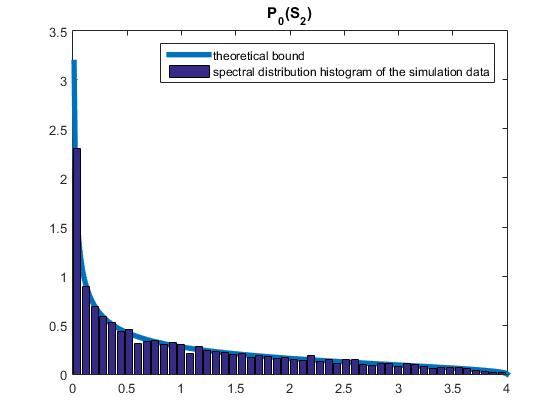}
 }\\
 \subfloat[Stable growth B $V_3$]{\label{fig1d}
 \includegraphics[width=0.3\textwidth]{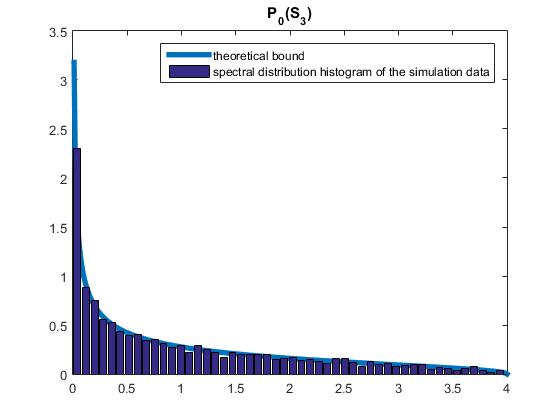}
 }
 \subfloat[Voltage collapse $V_4$]{\label{fig1e}
 \includegraphics[width=0.3\textwidth]{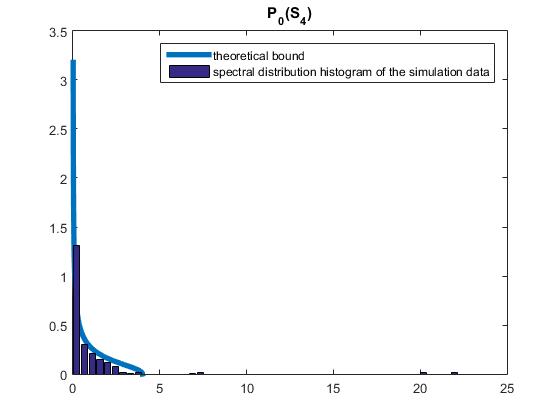}
 }
 \subfloat[White noises $V_5$]{\label{fig1f}
 \includegraphics[width=0.3\textwidth]{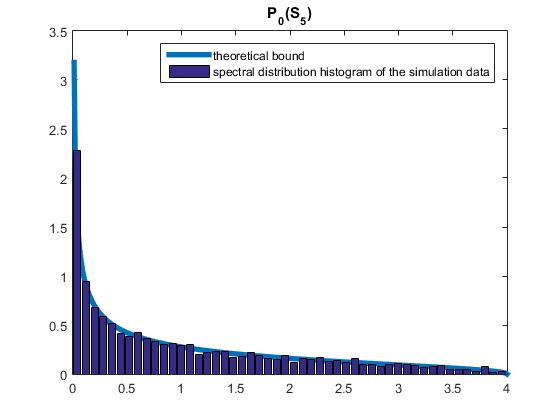}
 }
 \caption{}
 \label{fig3}
 \end{figure*}

1) Fig. 3(a)  and  (f) indicate that the histograms agree with M-P Law relatively well; there are only small noises in these two cross sections.

2) Fig. 3(b) and (e) show that there are several outliers and the outliers in (e) are lager than those in (b). It means that there is some anomaly in cross section $\textbf{C}_1$ and section $\textbf{C}_4$. Moreover,  the latter is more serious than the former.

The two results above agree with the previous simulation~\cite{He2015A} in our team. This agreement validates the code used here.

3) Then, we find that there is no outlier in Fig. 3(c) and (d), and the histogram curves in (c) and (d) are quite similar with those in (a) and (f). However, there are obvious ramp signals in their corresponding cross sections in fact.  It means that we can hardly distinguish the two steady load growths from each other, or even the steady growths from white noises.
This inspires us to conduct a new study of data fusion.

\subsection{Case 2: Detection with linear data fusion $P_1$}

We conduct data fusion between ${V_i}$ $(i \ge 1)$  and  the  reference matrix {$V_0$} through the proposed method. Note that {$S_i$} is generated from ${V_i}$ by the preprocess in III-C.
Then, we choose a multivariate linear polynomial  to conduct this simulation:
 \[{P_1}({S_0},{S_i}) = {S_0} + {S_i},~i=1,2,3,4,5.\]

 Simulation results are shown in Fig.~4. The theoretical bound is obtained through Theorem~\ref{th1}. The spectral distribution histogram of ${P_1}({S_0},{S_i})$ is plotted through Algorithm~\ref{alg1}.

\begin{figure*}
 \centering
 \subfloat[White noises $V_5$ \& $V_0$]{
 \includegraphics[width=0.3\textwidth]{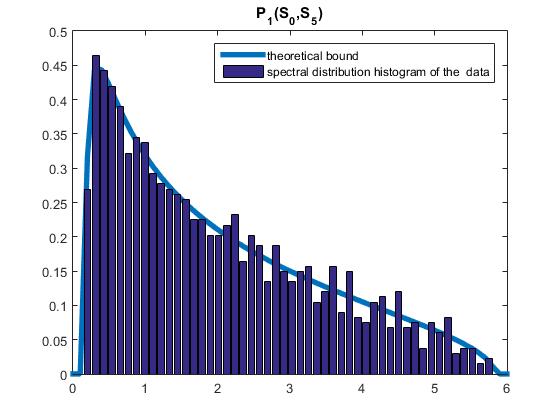}
 }
 \subfloat[Step signal $V_1$ \& $V_0$]{
 \includegraphics[width=0.3\textwidth]{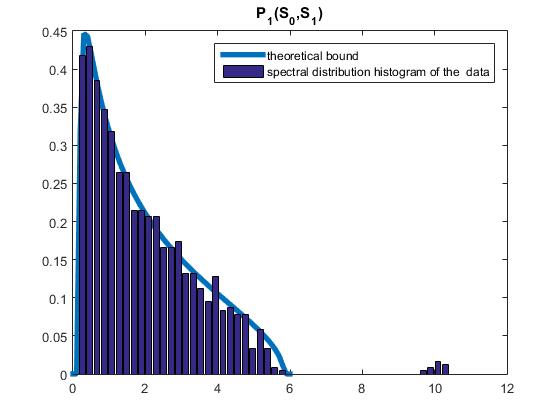}
 }
 \subfloat[Stable growth A $V_2$ \& $V_0$]{
 \includegraphics[width=0.3\textwidth]{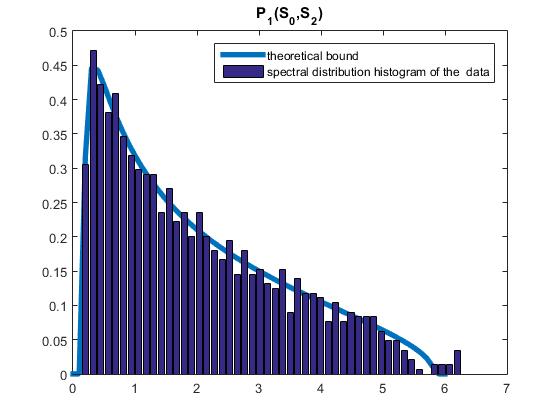}
 }\\
 \subfloat[Stable growth B $V_3$ \& $V_0$]{
 \includegraphics[width=0.3\textwidth]{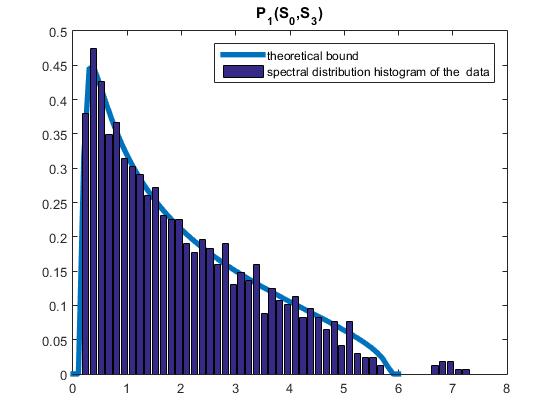}
 }
 \subfloat[Voltage collapse $V_4$ \& $V_0$]{
 \includegraphics[width=0.3\textwidth]{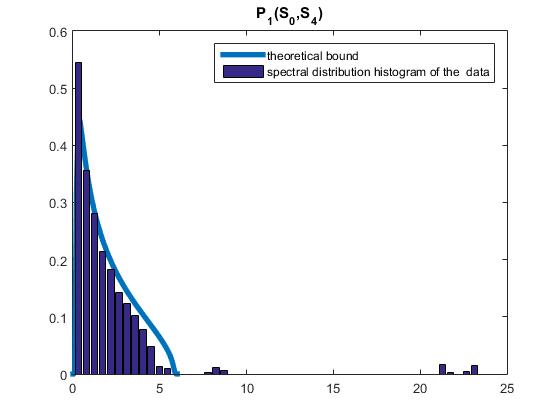}
 }
 \caption{Data fusion using multivariate linear polynomial $P_1$}
 \label{fig4}
 \end{figure*}

1)  For Fig. 4(a), the histogram matches the theoretical bound very well; for other figures there exist outliers.

2) Sort by the size of outliers: (a)$<$(c)$<$(d)$\ll$(b)$\ll$(e).

From result 1),  we obtain that this method detects the  signals from  white noise successfully. Result 2) means that the step signals and the ramp signals are distinguished by the size of outliers.
Furthermore,  in some way, the anomaly's influence on the grid can be estimated qualitatively by the size of outliers.

\subsection{Case 3: Detection with nonlinear data fusion $P_2$}
 Another multivariate polynomial, which is nonlinear, is chosen:
 \[{P_2}({S_0},{S_i}) = {S_0}{S_i} + {S_i}{S_0},i=1,2,3,4,5.\]

\begin{figure*}
 \centering
 \subfloat[White noises $V_5$ \& $V_0$]{
 \includegraphics[width=0.3\textwidth]{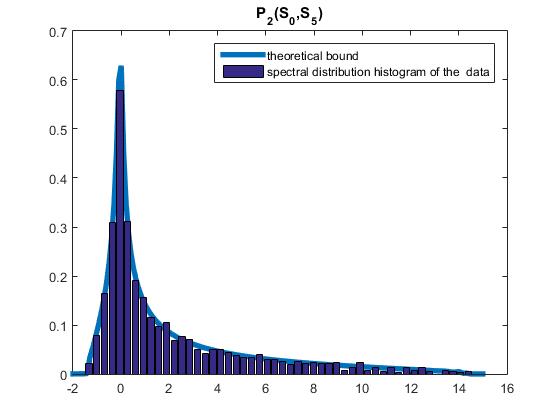}
 }

 \subfloat[Step signal $V_1$ \& $V_0$]{
 \includegraphics[width=0.3\textwidth]{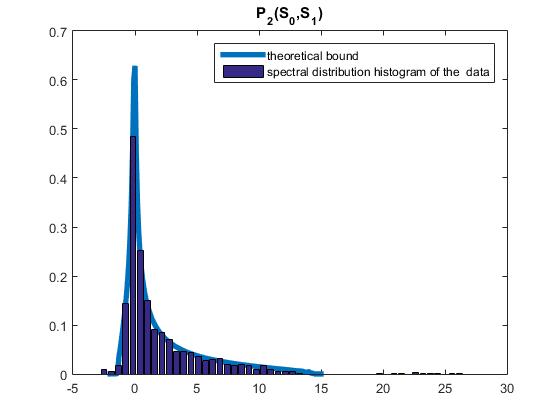}
 }
 \subfloat[Stable growth A $V_2$ \& $V_0$]{
 \includegraphics[width=0.3\textwidth]{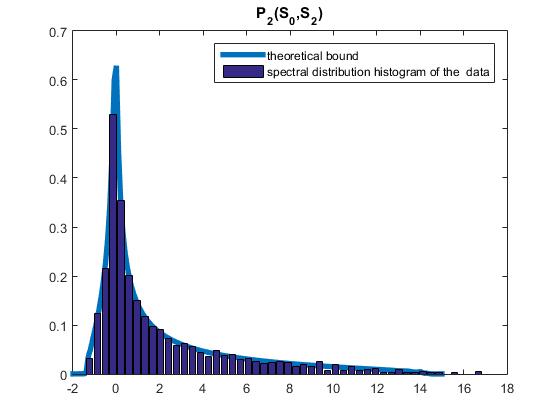}
 }\\
 \subfloat[Stable growth B $V_3$ \& $V_0$]{
 \includegraphics[width=0.3\textwidth]{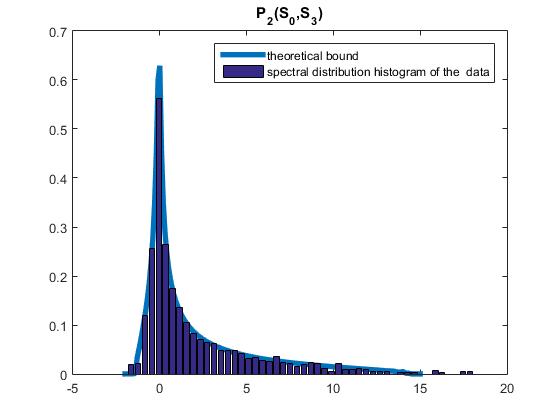}
 }
 \subfloat[Voltage collapse $V_4$ \& $V_0$]{
 \includegraphics[width=0.3\textwidth]{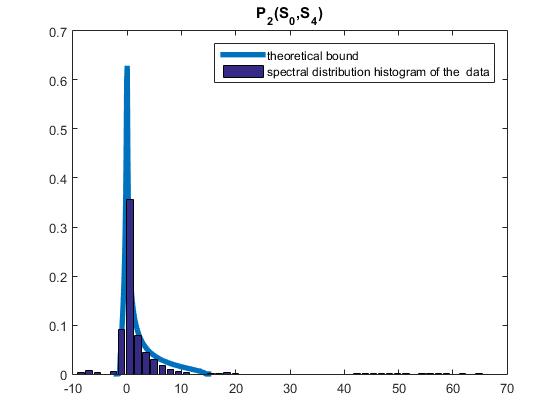}
 }
 \caption{Data fusion using multivariate nonlinear polynomial $P_2$}
 \label{fig5}
 \end{figure*}

Fig. \ref{fig5} shows the results. Both the process and analysis results of this simulation is similar to the subsection \textit{B} while the theoretical bound is obtained through the algorithm in Section~\ref{sect:AlgorithmFreeAdjointPolynomial}.

After comparing Fig.\ref{fig5} with Fig.\ref{fig4}, the histogram matches the theoretical bound more perfectly when the grid operates normally. Besides,  differences in the number or size of outliers are more obvious. Compared with the linearity, nonlinearity is more flexible in problem  modeling and closer to the nature of the objective things. In some sense,  some other multivariate nonlinear polynomials may be more effective for the power grid with special load  characteristics.
\section{CONCLUSION}
Based on the random matrix model, we can build statistical models using massive datasets across the power grid, and employ hypothesis testing for anomaly detection. The aim of this paper is to make the first attempt to apply the recent free probability result in extracting big data analytics,
in particular data fusion.  It is expected that this novel advanced analytical tool provides similar results to the previous work that is obtained using different algorithms. This also validates our current work. It is somewhat \textit{unexpected}, however, that some new findings are made possible by using this advanced mathematical algorithm only. For example, nonlinear polynomials of large random matrices yield better results than those of linear ones. It seems to imply that to conduct data mining (in depth) for massive datasets, such an advanced algorithm is essential. Although the work of this paper is still preliminary in terms of practical applications, the novel connection of the free probability with data fusion is reasonably successful.

Some questions remain unanswered. More applications of this theory are needed to evaluate the performance of this algorithm in different settings. In theory, this algorithm can handle both linear and nonlinear polynomials of large random matrices, up to an \textit{arbitrary order}.  In this paper,  we consider only the second order, though. How does the high order benefit the performance? How do we design the coefficients of the polynomials?

It is here worthwhile to notice that in a strict sense, free probability applies to infinite-dimensional random matrices. The convergence rate of the empirical spectral distribution to the asymptotic limits is a function of $1/N,$ where $N$ the node of the power grid in consideration. For $N=118,$ the accuracy is already sufficient to our practical problems.

Finally, our study is conducted in the settings of large power grid. Based on the spirit of our unified framework of using large random matrices in wireless network~\cite{qiu2012bookcogpp}, sensing~\cite{qiu2013bookcogsen} and smart grid~\cite{qiu2015smart}, we can explore settings for other fields. After all, the foundation of big data science can be firmly built on large random matrices~\cite{qiu2014foundation}.

  \begin{appendices}
        \section{The operator-valued setting}
        Let $\mathcal{A}$ be a unital algebra and $\mathcal{B}\subset \mathcal{A}$ be a subalgebra containing the unit. A linear map\[E:\mathcal{A}\to \mathcal{B}\] is a conditional expectation if
 \[E[b]=b  \ \quad for\;all \;b\in \mathcal{B}\]
 and
 \[E[{{b}_{1}}a{{b}_{2}}]={{b}_{1}}E[a]{{b}_{2}}  \ \quad for\;all \;a\in \mathcal{A} \ \;for\;all\; {{b}_{1}},{{b}_{2}}\in \mathcal{B}\]

 An operator-valued probability space consists of $\mathcal{B}\subset \mathcal{A}$ and a conditional expectation $E:\mathcal{A}\to \mathcal{B}$. Then, random variables ${{x}_{i}}\in \mathcal{A}(i\in I)$ are free with respect to $E$ (or free with amalgamation over $\mathcal{B}$ ) if $E[{{a}_{1}}\ldots {{a}_{n}}]=0$  whenever ${{a}_{i}}\in \mathcal{B}<{{x}_{j(i)}}>$ are polynomials in some ${{x}_{j(i)}}$ with coecients from $\mathcal{B}$ and $E[{{a}_{i}}]=0$ for all $i$ and $j(1)\ne j(2)\ne \cdots \ne j(n)$.

 In order to have some nice analytic behaviour, we will in the following assume
 that both $\mathcal{A}$ and $\mathcal{B}$ are ${{C}^{*}}$-algebras; $\mathcal{B}$ will usually be of the form $\mathcal{B}={{M}_{N}}(\mathbb{C})$, the $N\times N$-matrices. In such a setting and for $x={{x}^{*}}$, this $G$ is well-dened.

   \end{appendices}

\bibliographystyle{IEEEtran}
\bibliography{lznbib,helx}

\end{document}